\documentclass{article}

\usepackage[english]{babel}

\usepackage[a4paper,top=2cm,bottom=2cm,left=3cm,right=3cm,marginparwidth=1.75cm]{geometry}

\usepackage{amssymb}
\usepackage{siunitx}
\PassOptionsToPackage{hyphens}{url}\usepackage{hyperref}
\usepackage{cleveref}
\usepackage[utf8]{inputenc}
\usepackage[right]{lineno}
\usepackage{csquotes}
\usepackage{booktabs}
\usepackage{longtable}
\usepackage{adjustbox}
\usepackage{array}
\usepackage{url}
\usepackage{titlesec}
\usepackage{authblk}
\usepackage{xcolor} % Load the xcolor package for color options

\titleformat{\subsection}
  {\mdseries\itshape\large} % Medium series, italic shape, and large font size
  {\thesubsection}{1em}{} % Numbering, spacing, and the section title itself

\usepackage[english]{babel}
\usepackage[style=apa,backend=biber,natbib=true,maxcitenames=2,uniquelist=false]{biblatex}
\DeclareNameAlias{sortname}{family-given}
\DeclareNameAlias{default}{family-given}

\renewbibmacro{in:}{}
\DeclareFieldFormat[article]{title}{\mkbibquote{#1}\addcomma}
\DeclareFieldFormat[book]{title}{\mkbibemph{#1}\addcomma}
\DeclareFieldFormat[bookinbook]{title}{\mkbibemph{#1}\addcomma}
\DeclareFieldFormat[inbook]{title}{\mkbibquote{#1}\addcomma}
\DeclareFieldFormat[incollection]{title}{\mkbibquote{#1}\addcomma}
\DeclareFieldFormat[inproceedings]{title}{\mkbibquote{#1}\addcomma}
\DeclareFieldFormat[manual]{title}{\mkbibemph{#1}\addcomma}
\DeclareFieldFormat[misc]{title}{\mkbibemph{#1}\addcomma}
\DeclareFieldFormat[thesis]{title}{\mkbibemph{#1}\addcomma}
\DeclareFieldFormat[unpublished]{title}{\mkbibquote{#1}\addcomma}
\DeclareFieldFormat[patent]{title}{\mkbibemph{#1}\addcomma}
\DeclareFieldFormat[report]{title}{\mkbibemph{#1}\addcomma}
\DeclareFieldFormat[online]{title}{\mkbibquote{#1}\addcomma}
\DeclareFieldFormat[software]{title}{\mkbibemph{#1}\addcomma}
\DeclareFieldFormat[booklet]{title}{\mkbibemph{#1}\addcomma}
\DeclareFieldFormat[periodical]{title}{\mkbibemph{#1}\addcomma}
\DeclareFieldFormat[standard]{title}{\mkbibemph{#1}\addcomma}

\DeclareFieldFormat[article]{journaltitle}{\iffieldundef{shortjournal}{\mkbibemph{#1}\addcomma}{\mkbibemph{\printfield{shortjournal}}\addcomma}}
\DeclareFieldFormat{volume}{\bibstring{volume}~#1}
\DeclareFieldFormat{number}{\bibstring{number}~#1}

\DefineBibliographyStrings{english}{
  volume = {Vol.},
  number = {No.}
}

\renewbibmacro*{volume+number+eid}{%
  \printfield{volume}%
  \setunit*{\addspace}%
  \printfield{number}%
  \setunit{\addcomma\space}%
  \printfield{eid}}

\renewbibmacro*{journal+issuetitle}{%
  \usebibmacro{journal}%
  \setunit*{\addcomma\space}%
  \usebibmacro{volume+number+eid}%
  \setunit{\addcomma\space}%
  \usebibmacro{issue+date}}

\renewbibmacro*{publisher+location+date}{%
  \printlist{publisher}%
  \iflistundef{location}
    {\setunit*{\addcomma\space}}
    {\setunit*{\addcolon\space}}%
  \printlist{location}%
  \setunit*{\addcomma\space}%
  \usebibmacro{date}}

\DeclareCiteCommand{\cite}[\mkbibparens]
  {\usebibmacro{prenote}}
  {\usebibmacro{citeindex}%
   \usebibmacro{cite}}
  {\multicitedelim}
  {\usebibmacro{postnote}}

\renewbibmacro*{cite:labelyear+extrayear}{%
  \iffieldundef{labelyear}
    {}
    {\printtext[bibhyperref]{%
       \printfield{labelyear}%
       \printfield{extrayear}}}}

\renewbibmacro*{cite:labeldate+extradate}{%
  \iffieldundef{labelyear}
    {}
    {\printtext[bibhyperref]{%
       \printfield{labelyear}%
       \printfield{extradate}}}}

\AtEveryBibitem{
  \clearfield{month}
  \clearfield{day}
  \ifentrytype{book}{
    \clearlist{location}
  }{}
}

\DefineBibliographyStrings{english}{
  andothers = {\textit{et al.},}
}

\DeclareFieldFormat[article]{volume}{\bibstring{jourvol}\addnbspace #1}
\DeclareFieldFormat[article]{number}{\bibstring{number}\addnbspace #1}
\DeclareFieldFormat[article]{volume}{Vol. #1}
\DeclareFieldFormat[article]{number}{No. #1}
\DeclareFieldFormat{url}{\bibstring{available at}\addcolon\space\url{#1}}
\DeclareFieldFormat{urldate}{\mkbibparens{accessed \addspace#1}}

\DeclareFieldFormat{urldate}{%
  \mkbibparens{accessed\space%
    \thefield{urlday}\addspace%
    \mkbibmonth{\thefield{urlmonth}}\addspace%
    \thefield{urlyear}}}

\crefformat{figure}{#2Figure~#1#3}
\Crefformat{figure}{#2Figure~#1#3}
\crefformat{table}{#2Table~#1#3}
\Crefformat{table}{#2Table~#1#3}
\crefformat{section}{#2Section~#1#3}
\Crefformat{section}{#2Section~#1#3}

\author[1*]{J.  A. Martínez-Cadena}
\author[1]{J.  M. Sánchez-Cerritos}
\author[2]{A. Marin-Lopez}
\author[1]{J. Delgado-Fernández}
\author[3]{I. Fuentecilla-Carcamo}
\author[3]{E. Varela-Carlos}

\affil[1]{Departamento de Matemáticas, Universidad Autónoma Metropolitana-Iztapalapa, Iztapalapa, CDMX, 09340, México.}
\affil[2]{Departamento de Ingeniería de Procesos e Hidráulica. Universidad Autónoma Metropolitana-Iztapalapa, Iztapalapa, CDMX, 09340, México.}
\affil[3]{Facultad de Ciencias Físico-Matemáticas, Benemérita Universidad Autónoma de Puebla,  Puebla, 72000, México.}

\title{Wavelet analysis and forecast of pollutants \\ in  Puebla City, Mexico}
\date{} 

\begin{document}

\maketitle

\textbf{* Corresponding Author}. martinezcadenajuan@gmail.com 

\begin{abstract}
This article presents a detailed analysis of atmospheric pollutants in Puebla City, Mexico, based on data collected between 2016 and 2024. The research focuses on the daily variation of six major pollutants: ozone (O${}_3)$, particles smaller than 10 microns (PM${}_{10})$, particles smaller than 2.5 microns (PM${}_{2.5})$, sulfur dioxide (SO${}_2)$,  and nitrogen dioxide (NO${}_2)$. The Mann-Kendall test, Innovate Trend, and Wavelet Transform Analysis were applied to identify significant trends and seasonal patterns. The results indicate an increase in the levels of O${}_3$, SO${}_2$, and NO${}_2$, while the levels of PM${}_{10}$,   and PM${}_{2.5}$ have shown a decrease. The study also employs the Prophet Forecasting Model to predict PM${}_{2.5}$ and PM${}_{10}$ concentrations for the year 2022, demonstrating that the model's accuracy increases as the analysis extends over longer periods.
\\
\\
%add 6 keywords
\end{abstract}
%\linenumbers

\section{Introduction}
\label{sec:introduction}

According to the 2020 Population and Housing Census conducted by the Instituto Nacional de Estadística y Geografía (INEGI), the population of the municipality of Puebla was 1,692,181 inhabitants, making it the fourth largest city in the country after Mexico City, Monterrey and Guadalajara. Puebla is part of the Puebla-Tlaxcala metropolitan area, whose geographic position and relatively flat topology are key for trade throughout the country.  
 
\begin{figure}[ht]
 \centering
 \makebox[\textwidth][c]{\includegraphics[width=.6\textwidth]{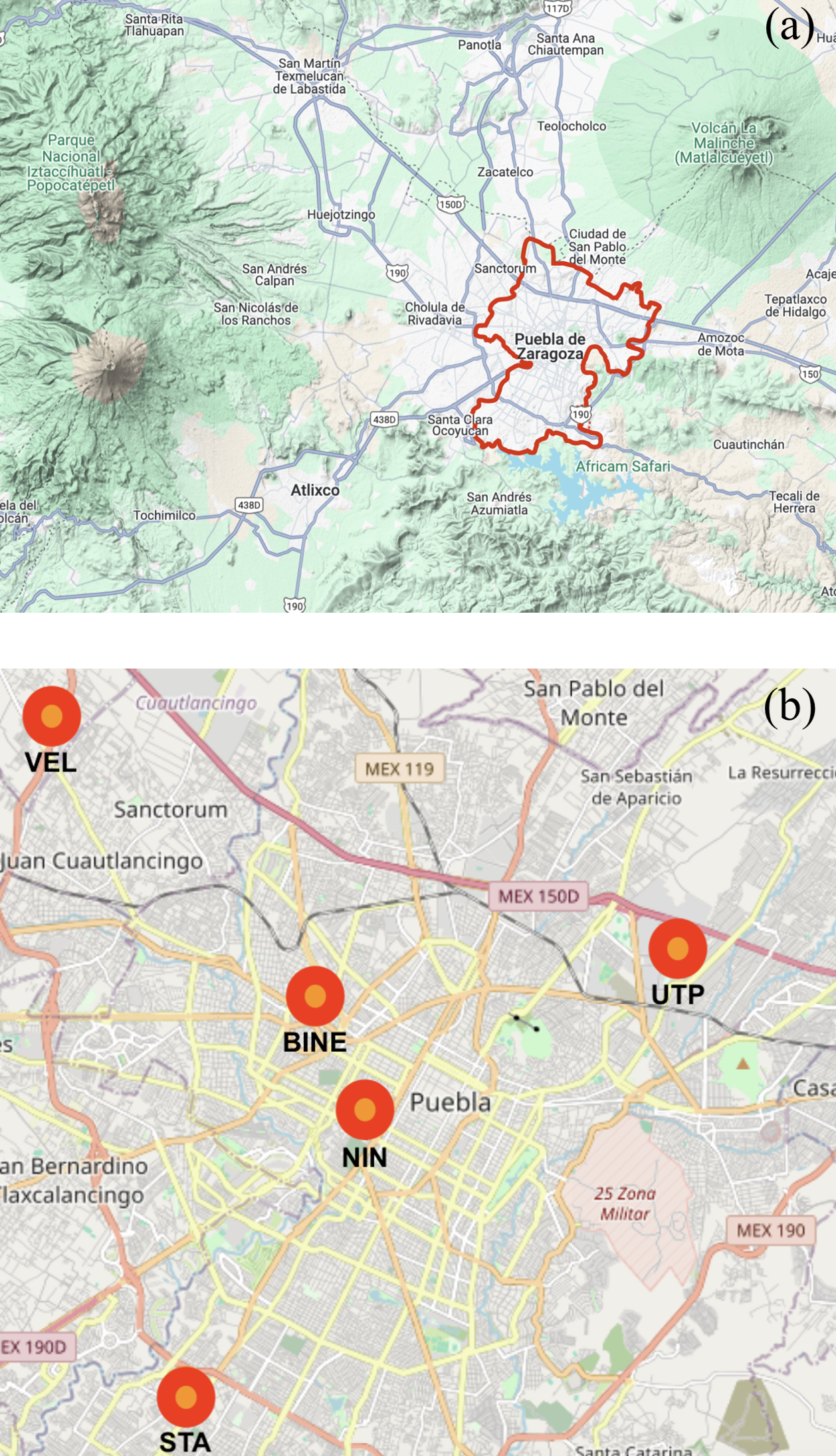}}%
 \caption{(a) Map of Puebla City and the surrounding areas.  (b) SINAICA atmospheric monitoring stations in Puebla city.}
 \label{stations}
\end{figure}

Puebla is located at an average altitude of 2,040 meters above sea level and is surrounded by mountain ranges of the Trans-Mexican Volcanic Belt.  To the west lie the Popocat\'epetl and Iztacc\'ihuatl volcanoes, to the north La Malinche,  and the east the Pico de Orizaba. The green area of Amozoc on the eastern edge of the urban area acts as a significant green lung for the city (Figure \ref{stations}-(a)).

The predominant climate in Puebla is temperate with varying humidity levels, with an average temperature of 16°C and summer as the rainiest season.  The  COVID-19 lockdown temporarily reduced the concentration of carbon monoxide, nitrogen dioxide, ozone, sulfur dioxide, and particulate matter smaller than 10 microns. However, Puebla City's air quality has been affected by a significant increase in pollutants since 2020 according to The Wheater Channel.

Mexican Official Standards NOMs describing maximum permissible limits for pollutants reflect recent information on health effects and air quality management. Two types of standards are used for air quality monitoring: environmental health NOMs that establish permissible limits for criteria pollutants, and technical NOMs that define measurement methods for criteria pollutants.

Over recent years, extensive research has shown how air pollution impacts health. Exposure to pollutants such as particulate matter and ozone is linked to increased death rates and hospitalizations due to respiratory and cardiovascular disease. Both short-term studies, examining daily pollution fluctuations, and long-term studies, following groups exposed to pollution over time, have found adverse effects even at very low exposure levels. Furthermore, it has been observed that these adverse effects occur even at very low levels of exposure (Brunekreef, et al.,  2002).

\section{Data}
\label{sec:data}
Air pollutants are monitored hourly by the National Air Quality Information System (SINAICA), which has five stations covering the metropolitan area.   In this study, the central region of Puebla was considered, using data from the Las Ninfas (NIN), Benemérito Instituto Normal del Estado (BINE), and Universidad Tecnológica de Puebla (UTP) stations for the period 2016-2024 (Figure \ref{stations}-(b)). Five air pollutants that are routinely measured were considered: O${}_3$, PM${}_{10}$, PM${}_{2.5}$, SO${}_2$ and NO${}_2$.  Data were obtained from the public access site \href{https://sinaica.inecc.gob.mx/index.php}{sinaica.inecc.gob.mx}. For each pollutant, there are an average of 3043 daily observations.

\section{Methodology}
\label{sec:methodology}
Several statistical and analytical methods were applied to assess pollutant trends and better understand their temporal and seasonal variations:

\subsection{Mann-Kendall Test}

The non-parametric Mann-Kendall (MK) test was used to identify significant trends in the pollutant data time series. (Gilbert, 1987). This methodology is widely used in trend analysis due to its robustness and simplicity. The test statistics $(S)$  of the series $p_1,p_2,p_3,…,p_n$ are estimated through equation (\ref{s}), where $n$ represents the total observation,  $p_k$ and $p_j $indicates the observed data of time $k$ and $j$,

\begin{equation}\label{s}
S=\sum_{j=1}^{n-1}\sum_{k=j+1}^n \ sgn   (p_k-p_j),
\end{equation} 

where the $sgn$ function is defined as:

\begin{equation}
\ sgn  (p_k-p_j)= \left\{ \begin{array}{lcc}
 1 & \mbox{if} & (p_k-p_j)>0 \\ \\ 0 & \mbox{if} & (p_k-p_j)=0 \\ \\ -1 & \mbox{if} & (p_k-p_j)<0 \end{array} \right.
\end{equation}

The variance of $S$ is estimated as follows:

\begin{equation}
VAR(S)=\frac{1}{18}\left[   n(n-1)(2n+5)-\sum_{j=1}^g T_j(T_j-1)(2T_j+5) \right]
\end{equation}

here,  $g$ is number of the linked group, $T_j$ refers to the extent of the $j$-$th$ linked number.  From $S$ and \textit{VAR(S)}, the standardized test measure $Z$ is calculated using equation (\ref{Z}).

\begin{equation}\label{Z}
Z = \left\{ \begin{array}{lcc}
\frac{S-1}{\sqrt{VAR(S)}} & \mbox{if} & S > 0 \\ 
\\ 0 & \mbox{if} & S=0 \\ 
\\ \frac{S+1}{\sqrt{VAR(S)}} & \mbox{if} & S<0 \end{array} \right.
\end{equation}

A positive value of $Z$ indicates increasing trends, while a negative value indicates decreasing trends in the time series data.

\subsection{Innovate Trend Analysis}

The Innovative Trend Analysis (ITA) method was introduced by Sen (Sen, 2012) to evaluate the trends of different time series data. This method does not consider normality, serial autocorrelation, and data length. It also takes outliers and autocorrelation into account. In this method, the given time series data is divided into two equal subseries and then each subseries is arranged in ascending order independently. The subseries are plotted against each other, the first half of the subseries on the $X$-axis and the second half on the $Y$-axis, to obtain a scatter diagram.  Finally, the 1:1 straight line is drawn, the data line above the 1:1 line indicates increasing-positive trends, and below the line indicates decreasing-negative trends (Sen, 2017). The indicator of ITA (ITA$_{ind}$) is estimated as follows:

\begin{equation}
\mbox{ITA}_{ind} = \frac{1}{n} \sum_{i=1}^n \frac{10(p_i-p_k)}{\overline{p_1}},
\end{equation}

where, $p_i$ and $p_k$ indicates the values of the first and second half, $n$ is the extent of each half,  and $\overline{p_1}$ is the mean of the first half.  The slope ($m_s$) of the series can be calculated using the following equation: 

\begin{equation}\label{ms}
m_s = \frac{2(\overline{p_1}-\overline{p_2})}{n},
\end{equation}

where $\overline{p_2}$ is the mean of the first half.

\subsection{Wavelet transform}

In recent years, wavelet transform theory has been widely investigated due to its application in numerous disciplines, including physics, numerical analysis, signal processing, probability, and statistics, among other areas. Its usefulness lies in its ability to approximate a function or signal while retaining spatial information using a set of functions known as wavelets. In addition, the wavelet transform is used to extract localized features of interest in a signal, leading to better data compression.

The continuous wavelet transform of a function $f(t)$ is defined as the convolution of $f(t)$  with an analyzer function $\psi(\sigma)$.  To be considered a wavelet the function must be localized in time as well as in frequency space, moreover, it must be an integrable function, that is, with zero mean (Farge,  1992).  It is also assumed that $\psi$ is normalized, that is,  $\int_{-\infty}^{\infty}\psi \psi^* d\sigma=1$, where $\psi^*$ is the complex conjugate.  For a scale $s$, location $u$ and time $t$,  the variable
\begin{equation}
\sigma=\frac{(u-t)}{s}
\end{equation}

can be seen as a dimensionless time scale.  For a given time function $f(t)$, the continuous wavelet transform is obtained by 

\begin{equation}
W(u,s)=\frac{1}{\sqrt{s}} \int_{-\infty}^{+\infty} \psi^*\left(   \frac{u-t}{s} \right) f(t) dt,
\end{equation}

The factor $1/\sqrt{s}$ is necessary to satisfy the normalization condition.

The set of basis functions is derived from a single function called the Wavelet Mother function, this will be the one that undergoes modifications to perform the analysis; it will be expanded or compressed, and translated along the signal. These modifications occur through the scaling and displacement parameters. In scaling, the wavelet is lengthened or compressed, which allows us to see both the details and the components of the signal globally. While the displacement refers to the path of the wavelet along the signal. 
 
\begin{figure}[ht]
 \centering
 \makebox[\textwidth][c]{\includegraphics[width=.7\textwidth]{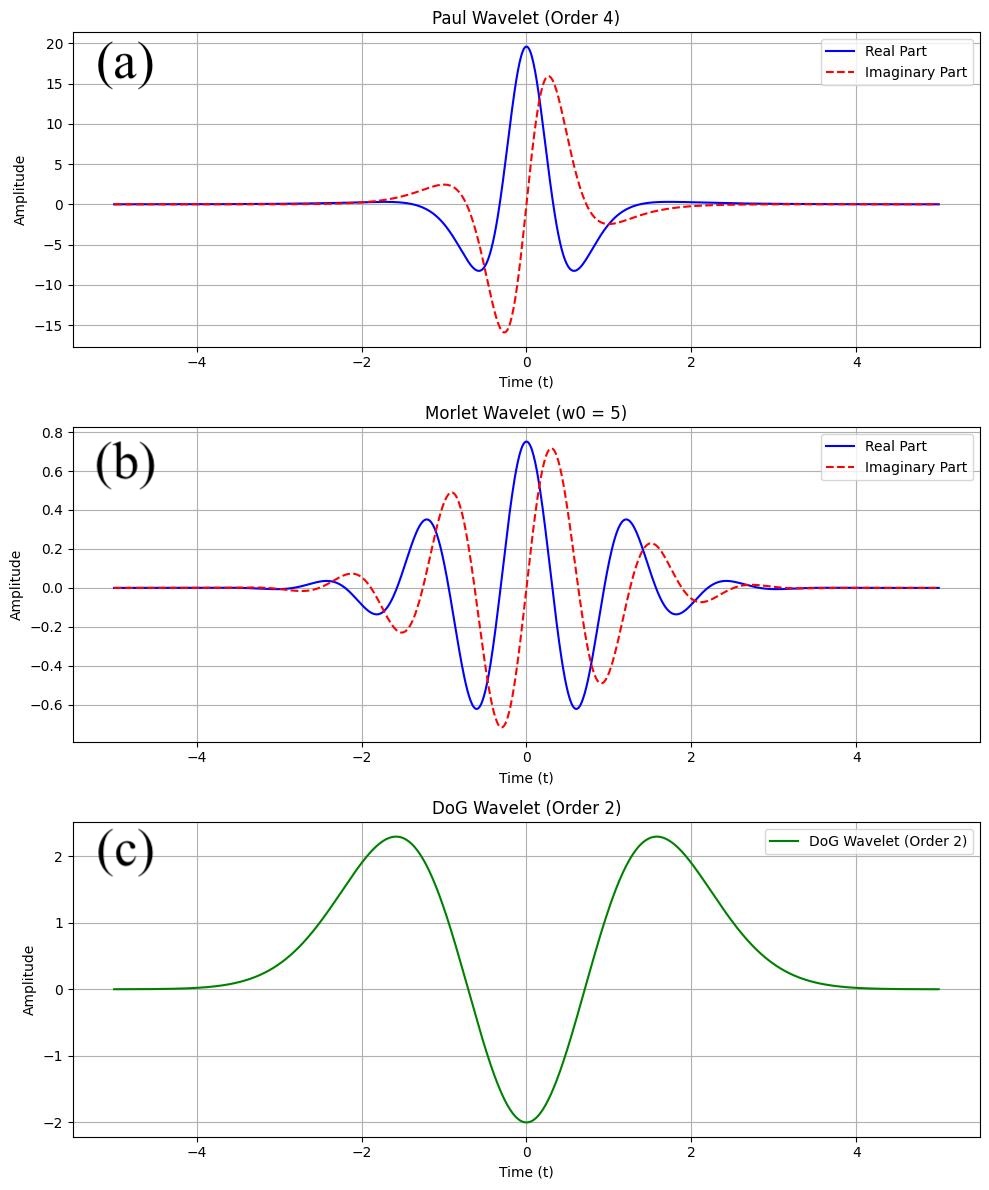}}%
 \caption{(a) Morlet wavelet, (b) Paul wavelet, and (c) DOG wavelet in the time domain}
 \label{mother}
\end{figure}

This paper concentrates on the three different mother wavelets given in (Torrence and Compo, 1998):

\begin{enumerate}
\item  The Morlet wavelet (Figure \ref{mother}-(a)):
\begin{equation}
\psi(\sigma)= \pi^{\frac{1}{4}} e^{i k \sigma} e^{-\frac{\sigma^2}{2}}
\end{equation}

\item The Paul wavelet (Figure \ref{mother}-(b)):
\begin{equation}
\psi(\sigma)= \frac{2^k i^k k!}{\sqrt{\pi (2\pi)!}} (1-i\sigma)^{-(k+1)}
\end{equation}

\item The Derivative of Gaussian (DOG) wavelet (Figure \ref{mother}-(c)):
\begin{equation}
\psi(\sigma)= \frac{(-1)^{k+1}}{\sqrt{\Gamma (k+\frac{1}{2})}} \frac{d^k}{d\sigma^k} \left(  e^{-\frac{\sigma^2}{2}}  \right)
\end{equation}

\end{enumerate}

The value of $k$ controls the number of oscillations present in the mother wavelet, and will therefore strongly influence the frequency and time resolution of the corresponding wavelet transform. The Morlet wavelet has a reasonably large number of oscillations, which will ensure good frequency resolution. The Paul wavelet has much fewer oscillations but is highly localized in time. This will give it very fine time resolution and at the same time reduced frequency resolution. The DOG wavelet has relatively few oscillations, over a much larger time domain. Note that both the Morlet and Paul wavelets are complex-valued, while the derivative of the Gaussian wavelet is real-valued. Wavelet analysis was used to extract localized features of interest in the signal, allowing for better data compression and the identification of seasonal cycles and variation patterns.

\subsection{Prophet Forecasting Model}

The Prophet Forecasting Model (PFM) is a regression model designed by the data science team of Facebook to handle time series data with daily observations. It is particularly useful for time series that display strong seasonal patterns such as air contaminant metrics for environmental analysis. The model decomposes the time series into three primary components: trend, seasonality, and holidays or special events, enabling a more flexible (PFM model incorporates some parameters that can be tuned) and interpretable forecasting. The PFM model is given by (Taylor, 2017)

\begin{equation}
y(t)=g(t)+s(t)+h(t)+\epsilon t
\end{equation}

Here, $y(t)$ represents the predicted value obtained from either a linear or logistic equation. The functions $g(t)$ and $s(t)$ capture the seasonality or time series patterns based on yearly, monthly, daily, or other periodic cycles, $h(t)$ accounts for outliers related to holidays, and $\epsilon(t)$ represents the random or unexpected error. In the PFM, change points are important parameters (they can be explicitly defined through the fitting scale) that represent moments in time when the underlying trend of a time series changes, that is, significant seasonal changes that impact the data, for example, holidays. To prevent overfitting and focusing on the most significant change points, PFM initially considers a large number of potential change points, and then an L1 regularization is applied to selectively narrow down to only a few key points. Due to the PFM characteristics to capture strong seasonal behavior, PFM has been successfully used to predict contaminant concentrations (Shen,  et al.,  2020) and (Hasnain,  et al.,  2022).  For the analysis at hand, training data corresponds to the first 6 years (2016-2021). After training, PFM was tested for the air quality of 2022.

\section{Results and Discussion}
\label{sec:ResultsandDiscussion}

\subsection{Descriptive Statistics}
\begin{figure}[ht]
 \centering
 \makebox[\textwidth][c]{\includegraphics[width=1\textwidth]{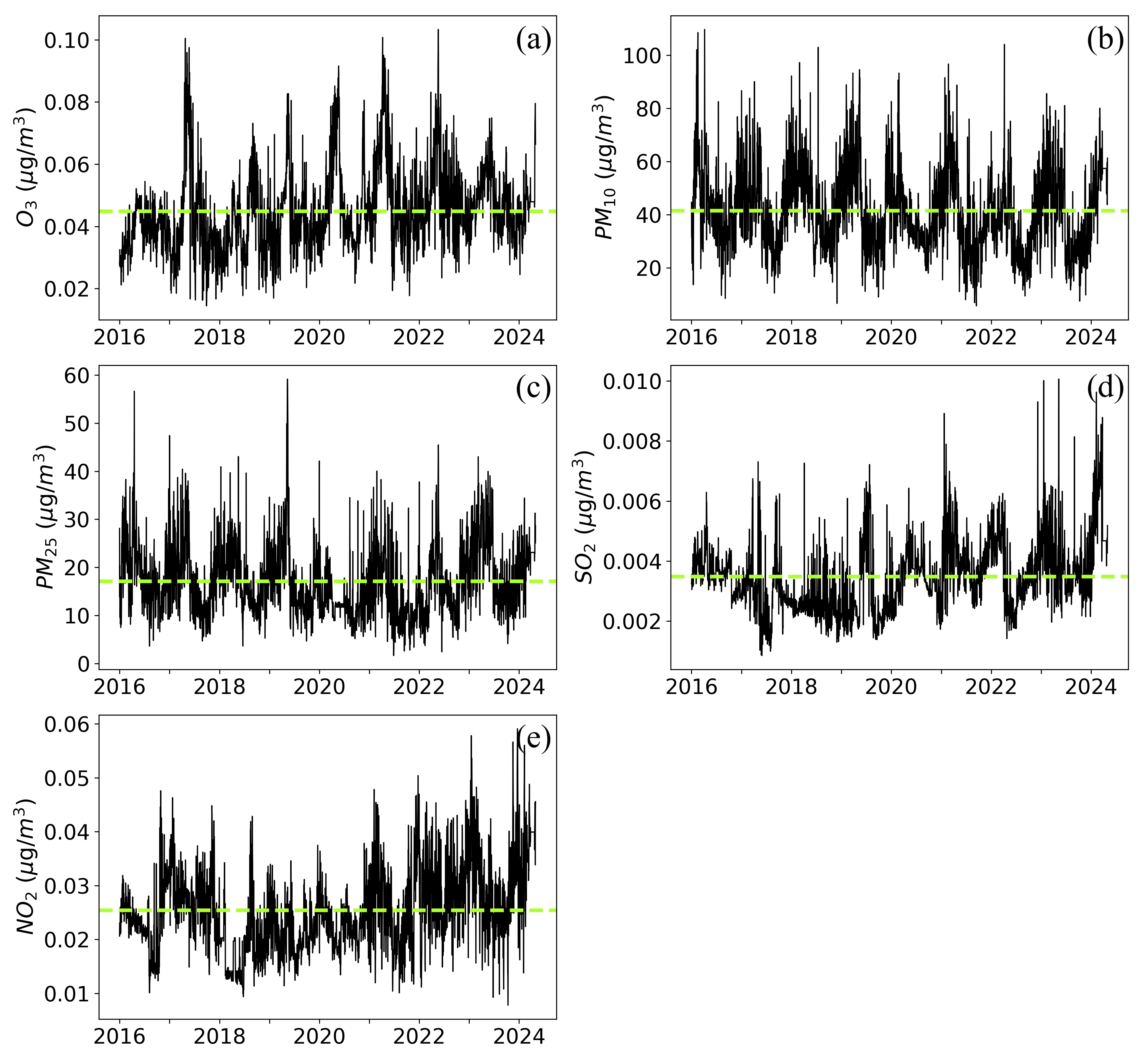}}%
 \caption{Daily variations of six air pollutants in Puebla City center region. (a) Ozone, (b) microparticles of size not smaller than 10 $\mu m$, (c) microparticles of size not smaller than 2.5 $\mu m$, (d) sulfur dioxide, and (e)  nitrogen dioxide. The vertical line denotes the mean of the time series.}
\label{series}
\end{figure}

Table \ref{sta} presents the summary statistics of the pollutants. Positive skewness is exhibited by the five series, indicating a trend to extremely high values of the pollutants rather than a trend showing reduced values. Fat tails are present in the five series as shown in terms of kurtosis and Shapiro-Wilk statistics. Figure \ref{series} exhibits the time series of the five pollutants in the period from January 2010 to June 2023. All pollutants showed an annual cycle. The maximum and minimum of the ozone cycle are found for the winter-spring and summer-autumn seasons, respectively. The blue line presents the average concentration.  The time series of ozone concentration in the City of Puebla indicates that ozone concentration shows sustained oscillations, these oscillations suggest stationarity, i.e., increases and decreases are experienced in a predictable seasonal cycle pattern. In addition, elevated ozone levels tend to form during periods of warm temperatures, so health-hazardous levels are common during the summer season. This level can increase during the warm season due to several interrelated factors. Ozone is mainly formed through chemical reactions involving sunlight, nitrogen oxides (NOx), and volatile organic compounds (VOCs).

\subsection{Trend Analysis}

\begin{figure}[ht]
 \centering
 \makebox[\textwidth][c]{\includegraphics[width=1\textwidth]{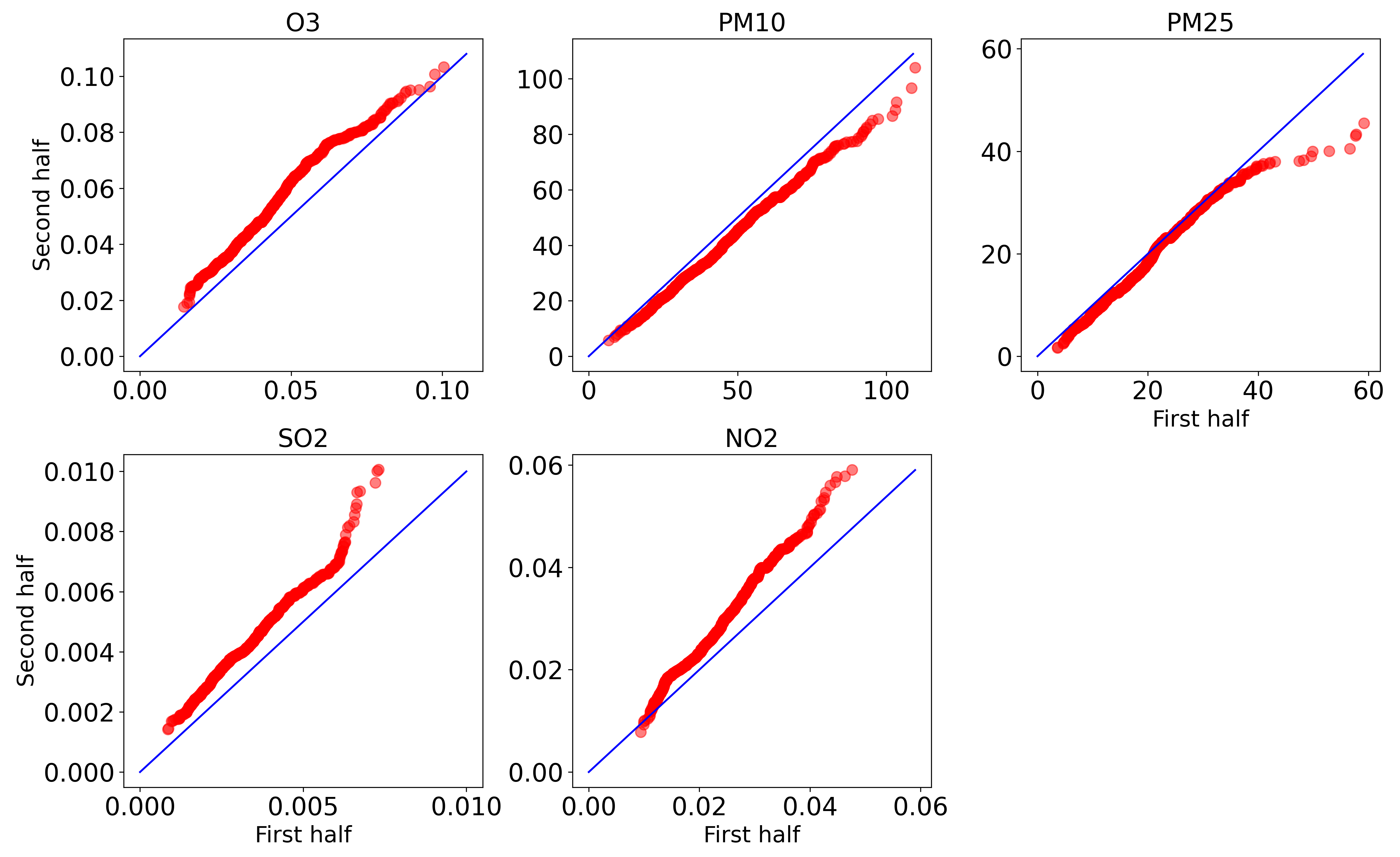}}%
\caption{Graphs obtained with the ITA method represent daily trends in pollutant concentration.}
\label{ITA}
\end{figure}

\begin{figure}[ht]
 \centering
 \makebox[\textwidth][c]{\includegraphics[width=1\textwidth]{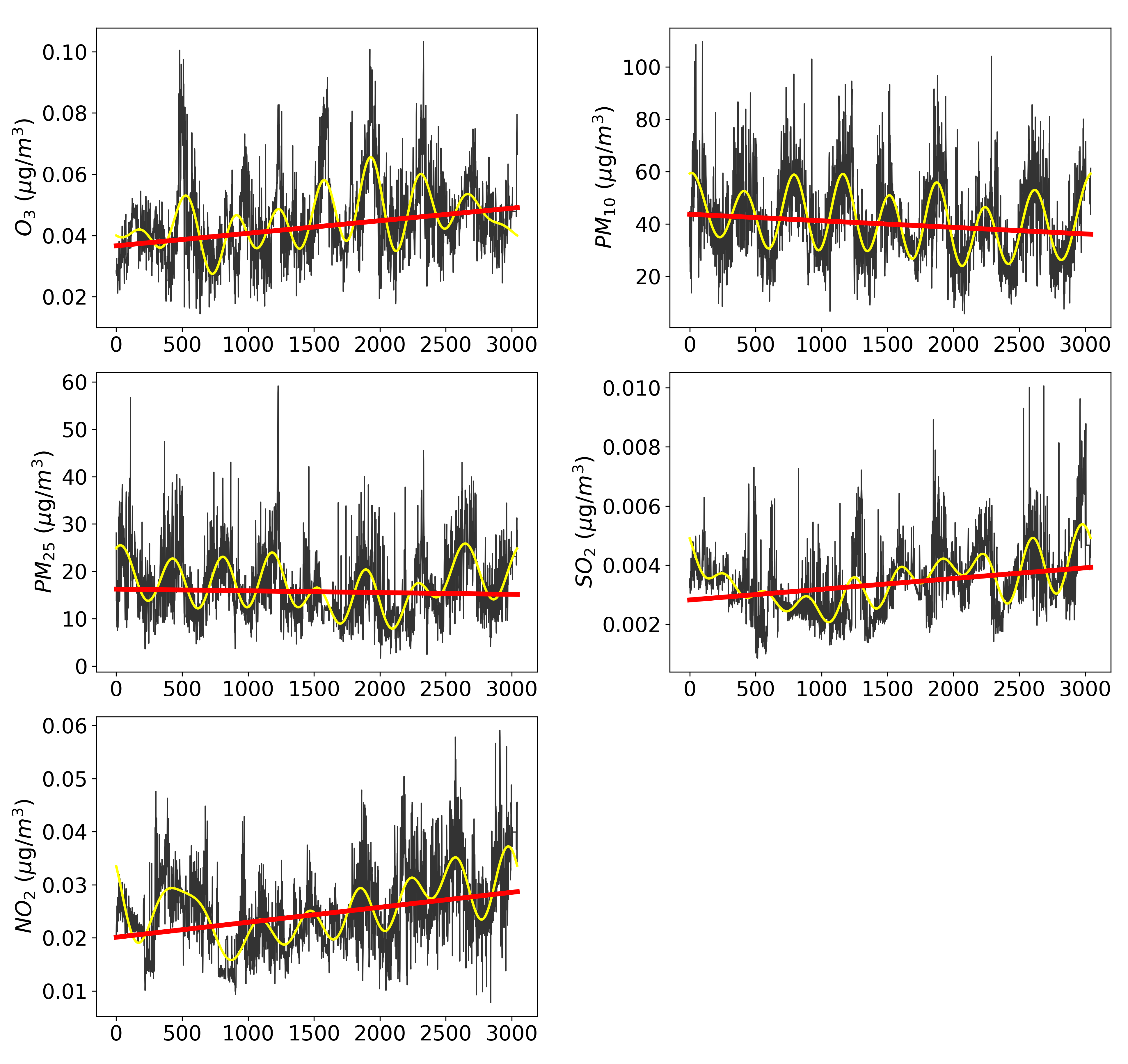}}%
 \caption{Trends in the pollutant series obtained with the slope of the ITA method.  The reading line corresponds to a low-pass filter with 10-day sampling used to highlight the long-term trend of the time series.}
 \label{ITAslope}
\end{figure}

Table \ref{mk} shows the trend obtained by the Mann-Kendall test. The positive value of $Z$ indicates increasing trends for O${}_3$,  SO${}_2$ and NO${}_2$, while a the negative value of $Z$ negative value indicates decreasing for for PM${}_{10}$, and PM${}_{2.5}$.  Using the ITA method, we can contrast this fact in the graphs (\ref{ITA}), where the pollutants with a negative trend appear below the 1:1 line and the pollutants with a positive trend appear above this line.  Finally,  using the equation (\ref{ms}), we can obtain a straight line that illustrates the trend of the time series data of the pollutants (Figure  \ref{ITAslope}), reaffirming the results given by the Mann-Kendall test.

\subsection{Wavelet Analysis}

\subsubsection{Ozone (O${}_3$)}

\begin{figure}[ht]
 \centering
 \makebox[\textwidth][c]{\includegraphics[width=1.2\textwidth]{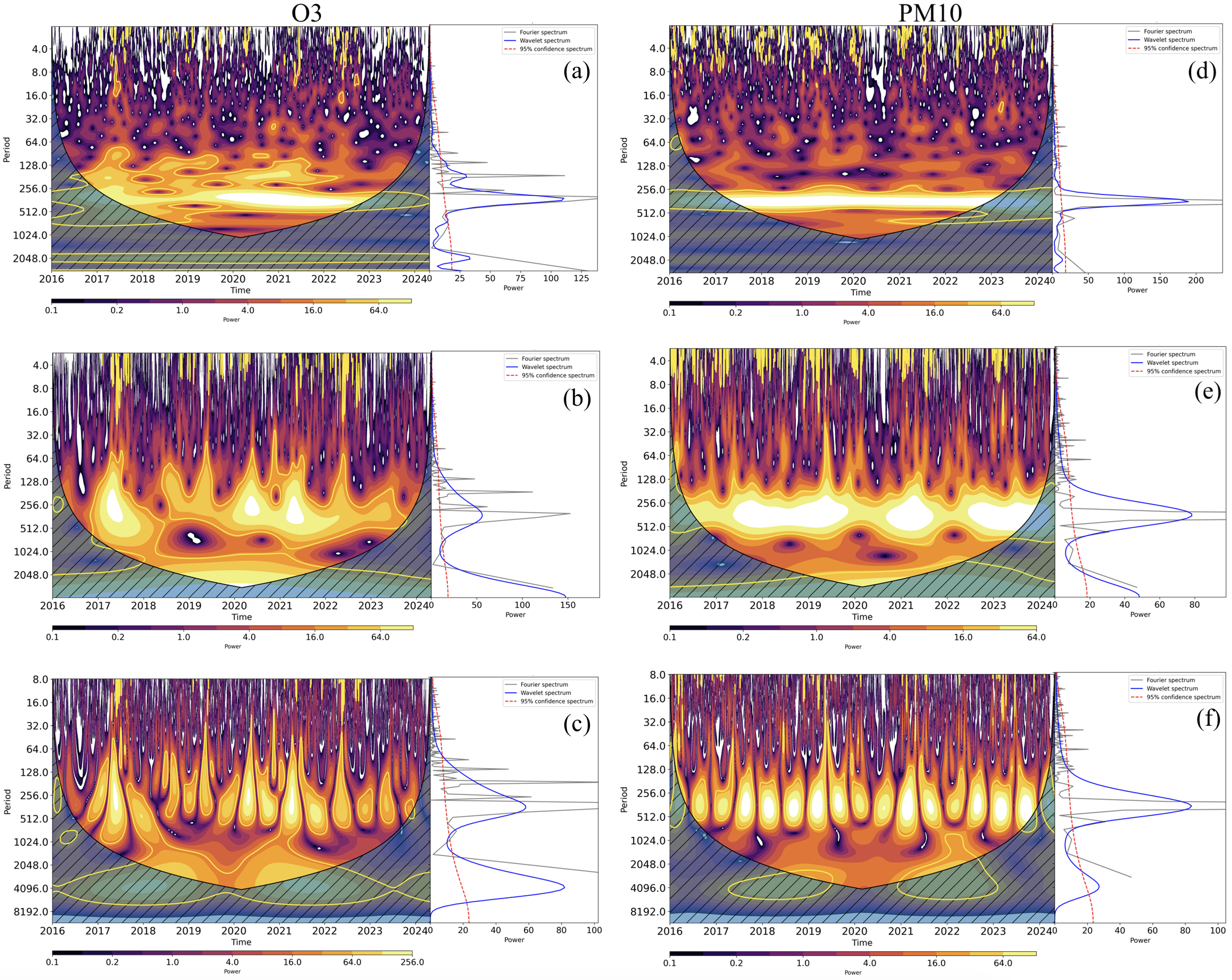}}%
 \caption{Scalograms of O${}_{3}$ and PM${}_{10}$, using the Morlet ((a), (d)), Paul ((b), (e)) and DOG ((c), (f)) wavelets.}
 \label{o3pm10}
\end{figure}

Figure \ref{o3pm10} presents the scalograms of the Morlet, Paul, and DOG wavelet spectra of O$_3$ and PM$_{10}$. In the case of O$_3$, the Morlet wavelet is used to generate the scalogram in Figure \ref{o3pm10}-(a). This complex wavelet is useful for detecting periodic patterns or dominant frequencies in the data. Here the x-axis is the period (in days) of the original signal corresponding to the wavelet scale on the y-axis, related to frequency. Wavelets with higher frequencies are plotted on lower scales and vice versa. The intensity or color indicates the magnitude of the transformation at a specific point in time and wavelet scale. The deeper the color or the higher the value, the higher the signal energy at that location. Therefore, the yellow contour encloses regions with confidence greater than 95\%. The largest yellow band is observed in the 360 days, indicating that we have a highly correlated signal, i.e. a pattern of strong influence. The hatched regions at each end indicate the ``cone of influence'' where edge effects become important. It is important to note that in the period from March 2020, the city of Puebla suffered a partial or total shutdown of factories, in addition to a lockdown due to the COVID-19 pandemic. However, this analysis showed that the O concentration did not decrease despite the measures imposed in response to the health crisis. A yellow coloration is even observed from 2020 to 2023 in approximately 360 days. In addition, high-energy bands can be visualized at smaller scales. The right side of Figure \ref{o3pm10}-(a) shows the Morlet wavelet spectrum and the Fourier spectrum. Here, the frequencies of the signals and their amplitude are shown. The peak of the dominant frequency is observed at 360 days. While the Fourier spectrum describes the frequencies present in a signal globally, the wavelet spectrum provides a time- and frequency-focused representation, making it easier to identify signal characteristics at various times and time scales.

Figure \ref{o3pm10}-(b) shows to case Paul's wavelet reflects a high-energy region on a 360-day scale. The cone of influence is narrower, suggesting that local perturbations have a more limited effect on the outcome of the analysis. Being a complex function that takes complex values instead of real values provides a more complete representation of the signal, as it includes more information about amplitude and phase. Here the Fourier spectrum shows two dominant peaks of energy, the predominant peak at about 360 days and a fainter peak at 130 days.

Figure \ref{o3pm10}-(c) presents the analysis performed on the time signal using the mother wavelet DOG. The most notable difference is the fine-scale structure using DOG. This might be because DOG is a real function that captures the positive and negative oscillations of the time series as separate peaks in the wavelet power. Morlet and Paul wavelets are complex functions containing more oscillations than DOG, and hence, the wavelet power combines positive and negative peaks into a single broad peak. Overall, the figures shown contain similar power features on the 360-day scale. Also, the DOG wavelet is narrower in space-time but broader in spectral space than the Morlet wavelet. Also, in Figure \ref{o3pm10}-(c), the peaks appear very sharp in the time direction but are more elongated in the scale direction.

\subsubsection{Particles smaller than 10 microns (PM${}_{10}$)}

Regarding the Morlet wavelet (\ref{o3pm10}-(d)), we can highlight that on a 360-day scale, there is a high-energy band throughout the entire period, we also observe significant autocorrelations at scales less than 8 days throughout the entire period. The use of this wavelet does not offer relevant information for this pollutant. The Paul wavelet shows a high-energy band throughout the entire period around 360 days, and this weakens at the beginning of 2020 and the beginning of 2022. This band shows a significant peak of autocorrelation from 32 days in mid-2019. In addition, we observe significant autocorrelations at scales less than 8 days throughout the entire period. Finally, in the scalogram corresponding to the DOG wavelet, we observe a high energy concentration in approximately semi-annual periods ranging from 200 to 500 days.

\subsubsection{Particles smaller than 2.5 microns (PM${}_{2.5})$}

\begin{figure}[ht]
 \centering
 \makebox[\textwidth][c]{\includegraphics[width=1.2\textwidth]{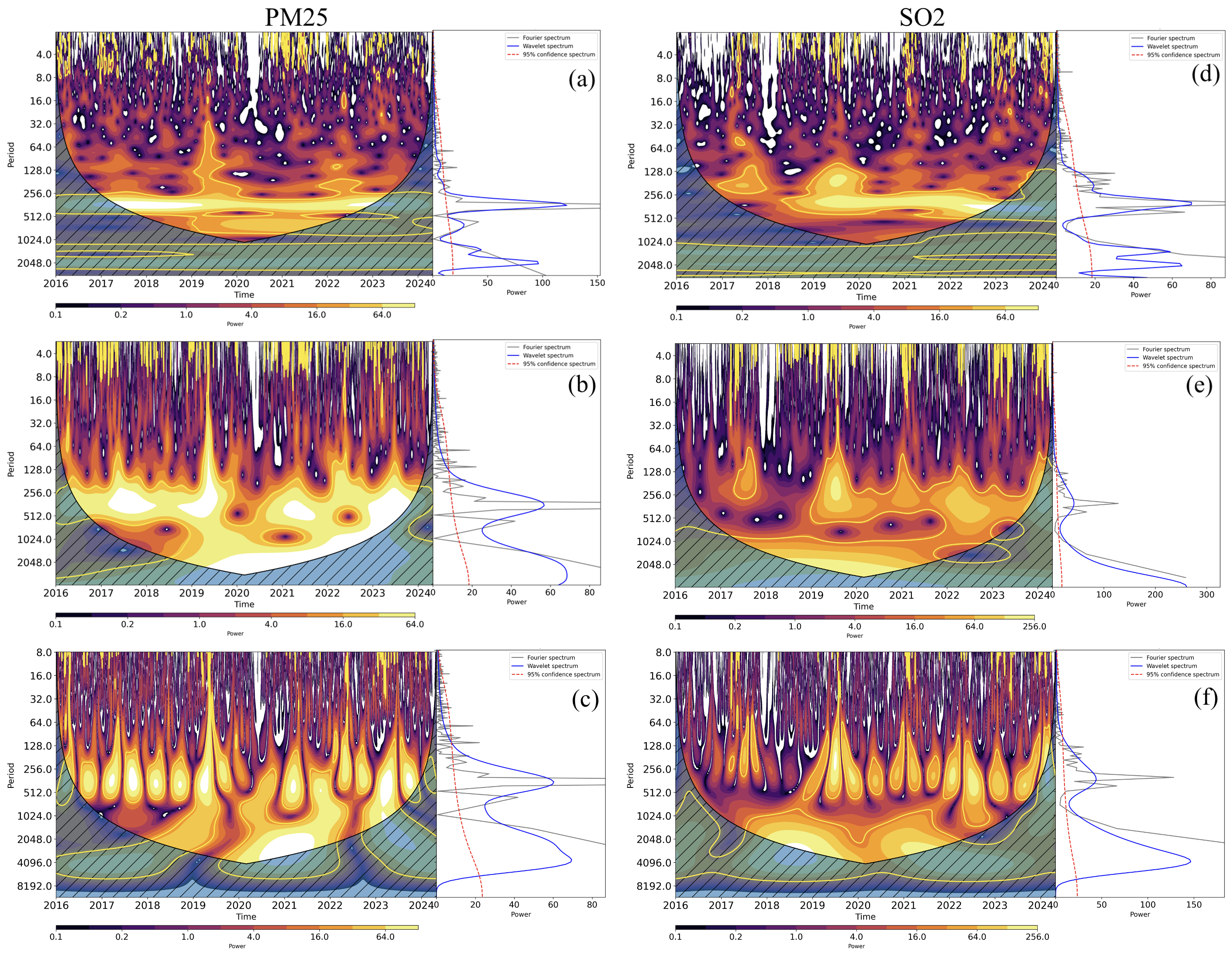}}
 \caption{Scalograms of PM${}_{2.5}$ and SO${}_{2}$, using the Morlet ((a), (d)), Paul ((b), (e)) and DOG ((c), (f)) wavelets.}
 \label{pm25so2}
\end{figure}

Regarding the Morlet wavelet (\ref{pm25so2}-(a)), we can highlight that on a 360-day scale, there is a thin band throughout the period, which decreases its power between 2020 and 2022. In addition, we observe significant autocorrelations at scales less than eight days throughout the entire period. The Paul wavelet shows a high-energy band around 360 days that begins to extend to very large scales starting in 2019. In the scalogram corresponding to the DOG wavelet, we observe a high-energy concentration in approximately semiannual periods ranging from 200 to 500 days and expanding to larger scales starting in 2019. Note that in all three scalograms, an important peak in signal energy appears in 2019.

\subsubsection{Sulfur dioxide (SO${}_2$)}

Figure \ref{pm25so2}-(a)(b)(c) shows the analysis performed for the SO${}_2$ pollutant with the Morlet, Paul, and DOG wavelet. Regarding the Morlet wavelet, we can highlight that on a 360-day scale, there is a high-energy band in time from 2019 to 2023; this band extends to the cone of influence, which is the area where the wavelet has a significant influence on the reconstruction of the original signal. Furthermore, it is interesting to note that we observe significant frequencies at both high and low scales. Recalling that higher scales represent lower frequency components and vice versa. The Paul wavelet shows some high energy zones between 2019 to 2020 and between 2021 to 2022. This energy extends from a scale of 64 to 512 days to the cone of influence. It is important to note that the signal also shows high energy on smaller scales. We notice that the 95\% confidence band of the spectrum is limited, this may be because the Paul wavelet is a real wavelet that is only based on magnitude and does not consider phase, which is crucial to understanding the complete structure of the signal. Finally, the wavelet dog is a wavelet-based on the difference between two Gaussians and is useful for detecting edges and local features. It is relatively simple and effective for identifying abrupt transitions. Here we see that the SO2 concentration increased on a scale of 360 days.  It is important to note that this wavelet indicates high energy at specific points.

\subsubsection{Nitrogen dioxide (NO${}_2$)}

\begin{figure}[ht]
 \centering
 \makebox[\textwidth][c]{\includegraphics[width=.6\textwidth]{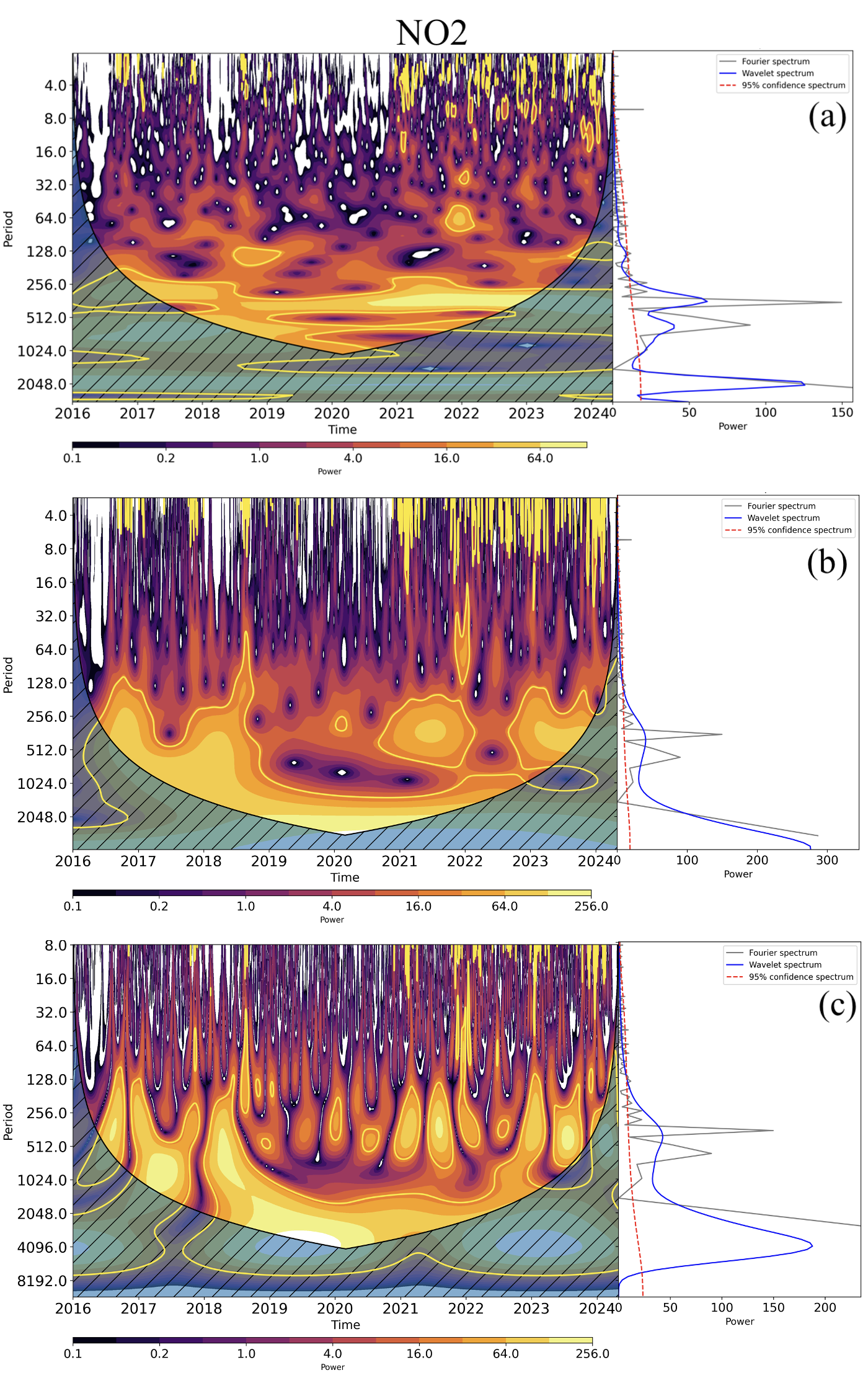}}%
 \caption{Scalogram of NO${}_{2}$, using the Morlet (a), Paul (b), and DOG (c) wavelets.}
 \label{no2}
\end{figure}

Figure \ref{no2} shows the analysis performed for the NO${}_2$ pollutant with the Morlet, Paul, and DOG wavelets. From here we can highlight that on a scale greater than 256 days, there is a high-energy band throughout the entire period, increasing the power of the signal from 2021 onwards. Figure \ref{no2}-(b) shows that the Paul wavelet indicates a high concentration of NO${}_2$ at small scales. Here we can see that there is no autocorrelation at scales less than 1,000 days during the periods 2019-2021 and 2022-2023. High concentration levels were present on scales greater than 16 days starting in 2021. For the DOG wavelet, we can observe the lack of autocorrelation at scales less than 1,000 days during the period 2019-2021. A significant peak of autocorrelation can be observed during the beginning of 2022 on all scales less than 512 days.

\subsection{PM${}_{10}$ and PM${}_{25}$ forecast}

\begin{figure}[ht]
 \centering
 \makebox[\textwidth][c]{\includegraphics[width=.7\textwidth]{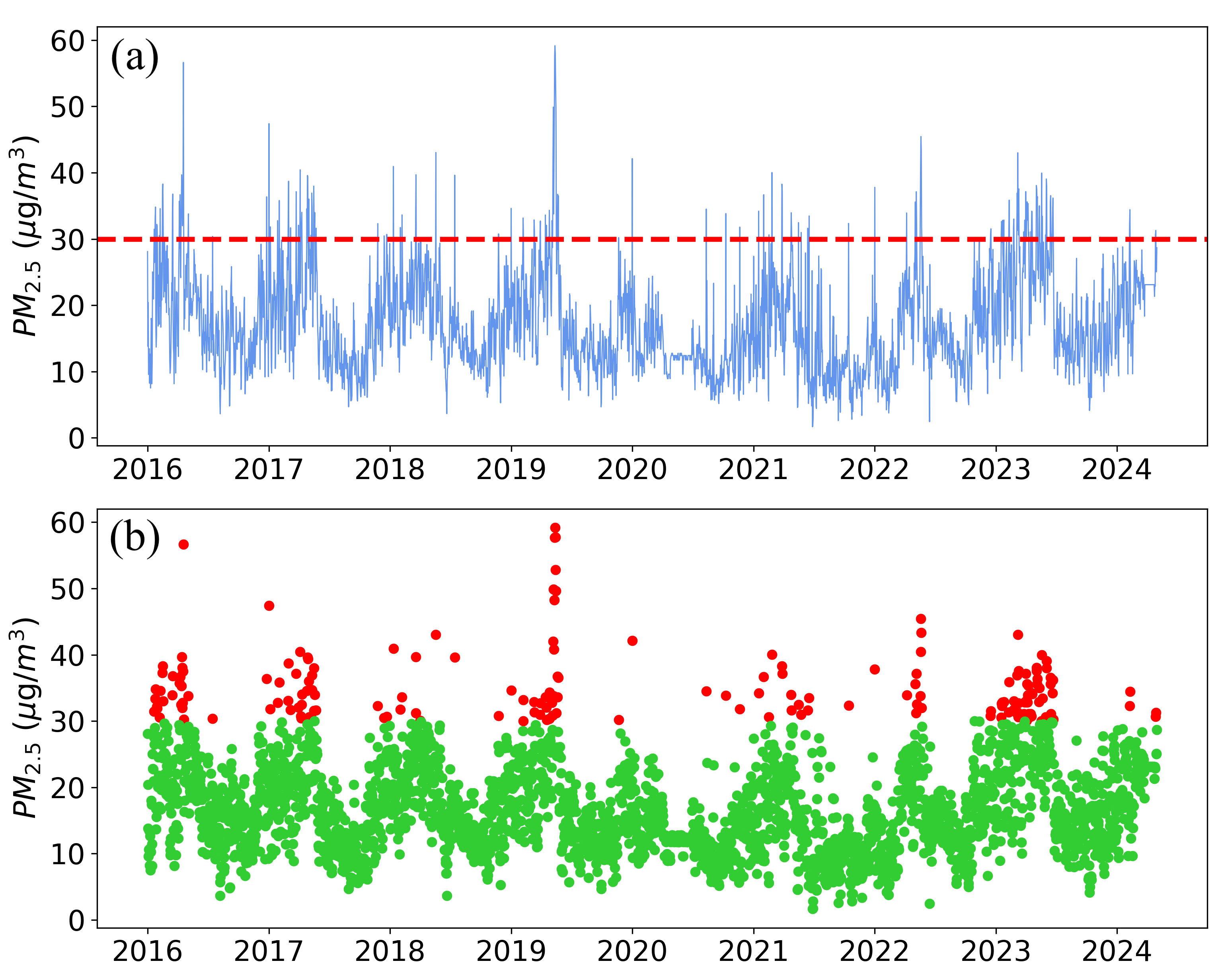}}%
 \caption{(a) PM${}_{2.5}$ time series. The horizontal dotted red line denotes the maximum daily exposure according to the NOM 025 guidelines. (b) The days in which the maximum daily exposure was exceeded are shown in red.}
 \label{pm25}
\end{figure}

\begin{figure}[ht]
 \centering
 \makebox[\textwidth][c]{\includegraphics[width=1\textwidth]{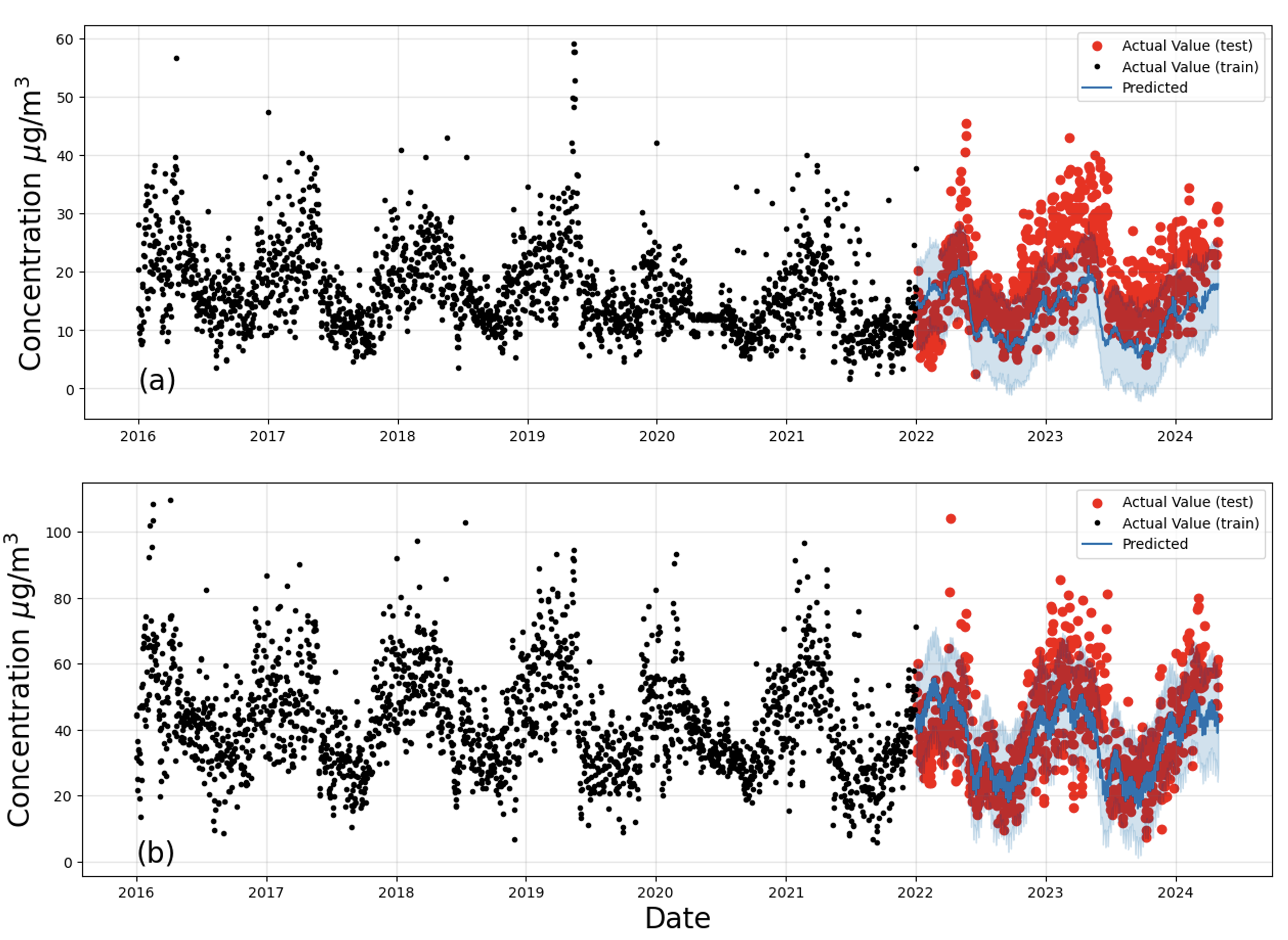}}%
 \caption{(a) PM${}_{2.5}$ and (b) PM${}_{10}$ forecasting in Puebla City. Scatter plots show actual values of pollutant concentration, and the PFM model is shown in blue.}
 \label{PM_PFM}
\end{figure}

Let us note that in the scalograms of the pollutants PM${}_{10}$, PM${}_{2.5}$ and SO${}_{2}$ in the previous section, an important peak in signal energy appears in 2019, this coincides with the fact that the highest concentrations of PM${}_{2.5}$ were recorded in May 2019 (Figure \ref{pm25}-(b)). In Mexico, the NORMA Oficial Mexicana NOM-025-SSA1-2014 (NOM 025) establishes that the concentration limit for PM${}_{2.5}$ is 30 $\mu g/m^3$, although the OMS recommends 25 $\mu g/m^3$. It is for this reason, that in this section,  our primary objective is to model concentrations of PM${}_{2.5}$,  and PM${}_{10}$ air-pollutants. We use PFM, developed by Facebook's data science team, to model both short-term and long-term trends. For our calculations, a linear model was used considering the official Mexican holidays and a COVID-19-related regressor that accounted for the impacts of the COVID-19 pandemic, specifically, the lockdown periods and restrictions implemented in Mexico as part of the semaphore epidemiológico (SE). SE system categorized regions into color-coded levels of restrictions -ranging from red (highest level of restrictions) to green (minimal restrictions)-based on the severity of COVID-19 spread. During the periods when Puebla was under stricter lockdown measures, such as when the semaphore was red or orange, there was a significant reduction in industrial activities, transportation, and public gatherings, all of which typically contribute to higher levels of air pollutants like PM${}_{10}$ and PM${}_{2.5}$. In Figure \ref{PM_PFM} we show the forecast model obtained from the test dataset (2016-2021) for (a) PM${}_{2.5}$ and (b) PM${}_{10}$. The model performance is determined using the correlation coefficient for the year 2022 (after training) and the results are given in table \ref{corr} for PM${}_{2.5}$ and PM${}_{10}$. As shown, the Prophet model exhibits weak performance in the short term, with a very low correlation between actual and predicted values for both PM${}_{2.5}$ and PM${}_{10}$. In this case, the model struggles to accurately capture the short-term fluctuations and dynamics of these pollutants, potentially due to complex and rapidly changing factors that the model can not account for. Over 6 months, the correlation increases to moderate levels, indicating that the model begins to capture more of the underlying trends in both PM${}_{2.5}$ and PM${}_{10}$ for long-term patterns.  For PM${}_{10}$, the model shows a strong ability of prediction over 9 to 12 months, with correlations in the range of 0.61 to 0.64. These values indicate that the model is effective at capturing long-term trends for PM${}_{10}$, possibly due to more stable or predictable patterns in this pollutant's behavior. For PM${}_{2.5}$, the correlation remains moderate at 0.40 over 9 months but decreases slightly to 0.30 over a full year, indicating that the PM${}_{2.5}$ model is less reliable over extended periods, possibly due to factors like variable sources or environmental conditions that are not fully modeled.

According to our calculations, the Prophet model performs better for PM${}_{10}$ than  PM${}_{25}$, over longer periods, implying that the utility of the model is more suitable for medium to long-term forecasting for PM${}_{10}$, while short-term predictions for both pollutants should be interpreted with caution due to weak correlations.

\begin{figure}[ht]
 \centering
 \makebox[\textwidth][c]{\includegraphics[width=1\textwidth]{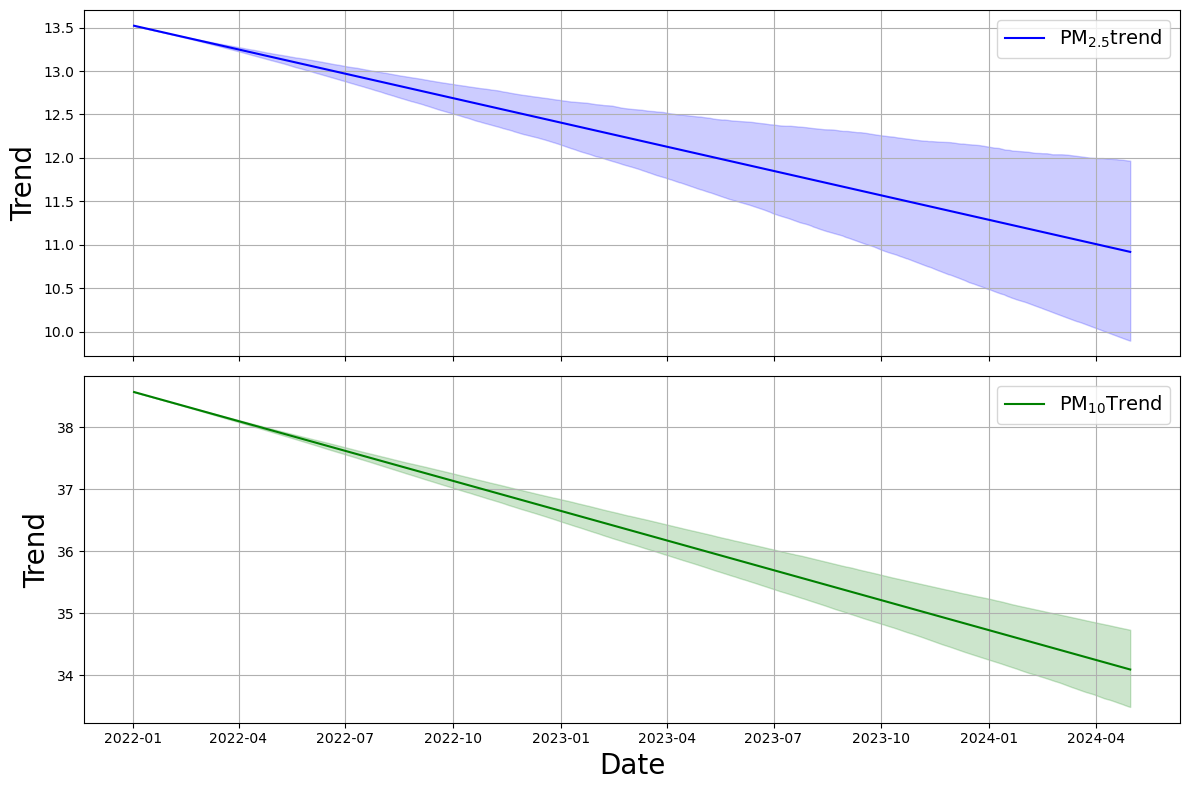}}%
 \caption{Decreasing trends for PM${}_{2.5}$ and PM${}_{10}$ using Porphet Forecasting model.}
 \label{PMtrends}
\end{figure}

Finally, we use the PFM to observe the underlying trend. The analysis of air quality data using the PM revealed significant trends for both PM${}_{2.5}$ and PM${}_{10}$ concentrations. After isolating the trend components of the model, the trend analysis showed a clear decreasing pattern over the study period. We computed the average rate of change for the test dataset starting from the year 2022, obtaining -0.0031 and -0.0053 for PM${}_{2.5}$ and PM${}_{10}$, respectively. In figure \ref{PMtrends}, the model's trend component indicates a consistent decline in PM${}_{2.5}$ and PM${}_{10}$  concentrations in accordance with the results of previous sections for PM${}_{10}$ and PM${}_{25}$.

\section{Conclusion}

The study highlights the dynamics of air pollution in Puebla City and its implications for public health and environmental policy. Key findings indicate significant increases in ozone (O${}_3)$, sulfur dioxide (SO${}_2)$, and nitrogen dioxide (NO${}_2)$ levels, while particulate matter (PM${}_{10}$ and PM${}_{2.5}$) and carbon monoxide (CO) levels have shown a decreasing trend. These changes can be attributed to factors such as the resumption of industrial activities, increased vehicular traffic, and the temporary effects of the COVID-19 lockdown.

To address these challenges, it is crucial to implement a dynamic and responsive air quality management strategy. This strategy should include precise identification of pollution sources through advanced spatial analysis, comprehensive health impact studies to quantify the burden of diseases linked to air pollution, and the integration of smart monitoring systems to enhance real-time data acquisition and response capabilities.

Policymakers must use these findings to develop and enforce stricter air quality regulations, promoting sustainable urban and industrial practices. Additionally, raising public awareness about the sources and health effects of air pollution can empower citizens to take personal measures to reduce their exposure and actively participate in air quality management processes. Further research and sustained policy efforts are essential to effectively address the complex environmental challenges posed by urban air pollution.

\vspace{1cm}

\textbf{Authors declared no conflict of interest} 

\medskip

\textbf{Generative AI tools were not used for the writing of the manuscript} 

\printbibliography

\newpage

\begin{table}
\begin{tabular}{lccccccc}
\toprule
Pollutant &	Min  & Mean  & 	Max & SD & 	Skewness & 	Kurtosis &	Shapiro-Wilk\\
 & ($\mu g/m^3$)	 &  ($\mu g/m^3$) &  ($\mu g/m^3$)	&  ($\mu g/m^3$) &  & 	 &	 \\

\midrule
O3 &	0.014 &	0.044 &	0.103 &	0.013	& 0.956	& 1.192 & 	0.948(0.00)\\
PM10 & 	5.739 & 	41.575	 & 109.680	& 15.498	& 0.597	& 0.372	& 0.978(0.00)\\
PM25	& 1.666	& 17.122	& 59.158	& 7.403	& 0.995	& 1.665	& 0.947(0.00)\\
SO2	 & 0.0008	& 0.003	& 0.010	& 0.001	& 0.987	& 1.843 & 	0.952(0.00)\\
%CO & 	0.326	& 1.195	 & 2.932	& 0.398	& 0.977	& 0.983 & 	0.940(0.00)\\
NO2	 & 0.007	& 0.025	& 0.059	 &  0.007	 & 0.657	& 0.446	& 0.973(0.00)\\
\bottomrule
\end{tabular}
\caption{Summary statistics of the five air pollutants.}
\label{sta}
\end{table}

\medskip

\begin{table}
\begin{center}

\begin{tabular}{lccccccc}
\toprule
Pollutant &	trend  & Z  & 	$\tau$ & S & 	Var(S) \\

\midrule
O3 &	increasing  &	17.59 (0.00)&	0.21 &	984559	& 3.132395$e^9$	\\

PM10 & 	decreasing  & -7.79 (0.00)&	-0.09	 & -436067 &		3.132397$e^9$\\

PM25	& decreasing 	& -2.53 (0.01)&  -0.03 & -141998	& 3.132386$e^9$ \\

SO2	 & increasing 	&15.53 (0.00)&0.18	& 869403	& 3.132395$e^9$	\\

%CO & 	decreasing &-18.73	(0.00)& -0.22	 & -1048488	& 3.132129$e^9$ \\

NO2	 & increasing 	& 17.23 (0.00)& 0.20	& 964650 &  3.132391$e^9$  \\
\bottomrule
\end{tabular}
\caption{Summary Mann-Kendall test of the six air pollutants.}
\label{mk}
\end{center}
\end{table}

\medskip

\begin{table}
\centering
\begin{tabular}{lcc}
\toprule
Time &	PM${}_{25}$  & PM${}_{10}$ \\

\midrule
30 days &	0.03 &	0.05 \\
90 days & 	0.13 & 	0.11\\
180 days	& 0.40	& 0.42\\
270 days	 & 0.40	& 0.64\\
360 days & 	0.30	& 0.61\\
\bottomrule
\end{tabular}
\caption{Correlation coefficient,  R, for PM${}_{25}$ and PM${}_{10}$ model.}
\label{corr}
\end{table}

\newpage

\newpage

\newpage

\newpage

\newpage

\newpage

\newpage

\newpage

\newpage

\newpage

\newpage

\end{document}